\documentclass[11pt,a4paper,onecolumn, unpublished]{quantumarticle}
\pdfoutput=1 
\usepackage[T1]{fontenc}

\usepackage{tikz}
\usetikzlibrary{positioning}
\usetikzlibrary{calc}
\usetikzlibrary{arrows}
\usepackage{mdwlist}
\usepackage{mathtools}
\usepackage{physics}
\usepackage{array}
\usepackage[numbers, sort&compress]{natbib}
\usetikzlibrary{decorations.pathmorphing}
\usepackage{mathrsfs}

\tikzset{snake it/.style={decorate, decoration=snake}}

\usetikzlibrary{decorations.pathreplacing,decorations.markings}
\tikzset{late arrow/.style={postaction={decorate},
  decoration={markings,mark=at position 0.85 with {\arrow{stealth}}}}}

\usepackage[dvipsnames]{xcolor}
\usepackage[colorlinks=true,
            linkcolor=blue,
            citecolor=blue,
            urlcolor=blue]{hyperref}
\usepackage[capitalize,nameinlink]{cleveref}
\usepackage{float}		
\usepackage{graphicx}		
\usepackage{subcaption}

\DeclareMathOperator{\opt}{\text{opt}}

\newcommand{\id}{\mathcal{I}}

\newcommand{\SWAP}{\mathrm{SWAP}}

\tikzset{
    >=stealth',
    punkt/.style={
           rectangle,
           rounded corners,
           draw=black, very thick,
           text width=6.5em,
           minimum height=2em,
           text centered},
    pil/.style={
           ->,
           thick,
           shorten <=2pt,
           shorten >=2pt,},
  on each segment/.style={
    decorate,
    decoration={
      show path construction,
      moveto code={},
      lineto code={
        \path [#1]
        (\tikzinputsegmentfirst) -- (\tikzinputsegmentlast);
      },
      curveto code={
        \path [#1] (\tikzinputsegmentfirst)
        .. controls
        (\tikzinputsegmentsupporta) and (\tikzinputsegmentsupportb)
        ..
        (\tikzinputsegmentlast);
      },
      closepath code={
        \path [#1]
        (\tikzinputsegmentfirst) -- (\tikzinputsegmentlast);
      },
    },
  },
  mid arrow/.style={postaction={decorate,decoration={
        markings,
        mark=at position .5 with {\arrow[#1]{stealth'}}
      }}}
}

\newtheorem{theorem}{Theorem}

\newtheorem{definition}[theorem]{Definition}

\newtheorem{lemma}[theorem]{Lemma}

\newtheorem{remark}[theorem]{Remark}

\newenvironment{proof}[1][Proof]{\noindent\textbf{#1.}}{\ \rule{0.5em}{0.5em}}

\usepackage{thmtools}

\begin{document} 

\title{Scrambling leads to entanglement cost in NLQC}

\author[1]{Atsuya Hasegawa}
\email{atsuya.hasegawa@math.nagoya-u.ac.jp}
\orcid{0009-0006-7151-2793}

\author[2,3]{Alex May}
\email{amay@perimeterinstitute.ca}
\orcid{0000-0002-4030-5410}

\affiliation[1]{Graduate School of Mathematics, Nagoya University}
\affiliation[2]{Institute for Quantum Computing, Waterloo, Ontario}
\affiliation[3]{Perimeter Institute for Theoretical Physics, Waterloo, Ontario}

\begin{abstract}
In a non-local quantum computation (NLQC), shared entanglement along with a single simultaneous round of quantum communication is used to implement a joint channel. 
NLQC has many applications, most of which benefit from an understanding of lower bounds on the entanglement needed to implement a given channel as an NLQC. 
However, the properties of a channel that lead to it having an entanglement cost in NLQC have remained unclear, with most lower bounds only applying to specially constructed examples.
Here, we give a new lower bound technique for unitaries based on scrambling: we find that a unitary has zero entanglement cost if and only if it does not scramble. 
Further, we find that the tripartite mutual information, an entropic object that quantifies the degree of scrambling, provides a quantitative lower bound on entanglement cost in NLQC. 
This is the first known lower bound on entanglement cost for unitaries in NLQC which is faithful. 
Our results also relate scrambling and NLQC for the first time.
Leveraging the faithful scrambling based bound, we then show that the controllable correlation, an earlier studied quantity, is also a faithful lower bound, resolving an earlier conjecture of Cleve and May.
\end{abstract}

\maketitle

\tableofcontents

\section{Introduction} 

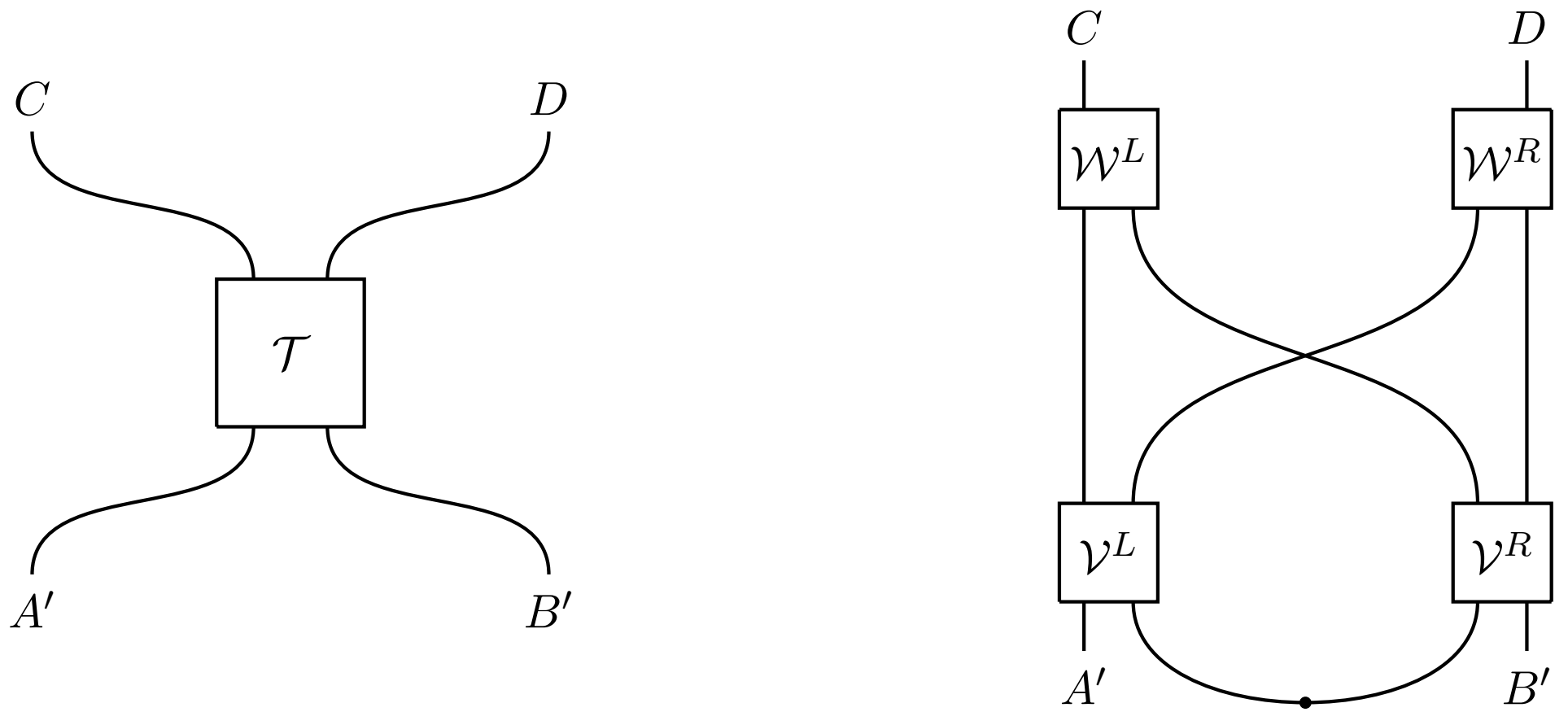
\begin{figure*}
    \centering
    \begin{subfigure}{0.45\textwidth}
    \centering
    \begin{tikzpicture}[scale=0.6]
    
    \draw[thick] (-1,-1) -- (-1,1) -- (1,1) -- (1,-1) -- (-1,-1);
    
    \draw[thick] (-3.5,-3) to [out=90,in=-90] (-0.5,-1);
    \node[below] at (-3.5,-3) {$A'$};
    \draw[thick] (3.5,-3) to [out=90,in=-90] (0.5,-1);
    \node[below] at (3.5,-3) {$B'$};
    
    \draw[thick] (0.5,1) to [out=90,in=-90] (3.5,3);
    \node[above] at (3.5,3) {$D$};
    \draw[thick] (-0.5,1) to [out=90,in=-90] (-3.5,3);
    \node[above] at (-3.5,3) {$C$};
    
    \node at (0,0) {$\mathcal{T}$};

    \node at (0,-5) {$ $};
    
    \end{tikzpicture}
    \caption{}
    \label{fig:local}
    \end{subfigure}
    \hfill
    \begin{subfigure}{0.45\textwidth}
    \centering
    \begin{tikzpicture}[scale=0.4]
    
    \draw[thick] (-5,-5) -- (-5,-3) -- (-3,-3) -- (-3,-5) -- (-5,-5);
    \node at (-4,-4) {$\mathcal{V}^L$};
    
    \draw[thick] (5,-5) -- (5,-3) -- (3,-3) -- (3,-5) -- (5,-5);
    \node at (4,-4) {$\mathcal{V}^R$};
    
    \draw[thick] (5,5) -- (5,3) -- (3,3) -- (3,5) -- (5,5);
    \node at (4,4) {$\mathcal{W}^R$};
    
    \draw[thick] (-5,5) -- (-5,3) -- (-3,3) -- (-3,5) -- (-5,5);
    \node at (-4,4) {$\mathcal{W}^L$};
    
    \draw[thick] (-4.5,-3) -- (-4.5,3);
    
    \draw[thick] (4.5,-3) -- (4.5,3);
    
    \draw[thick] (-3.5,-3) to [out=90,in=-90] (3.5,3);
    
    \draw[thick] (3.5,-3) to [out=90,in=-90] (-3.5,3);
    
    \draw[thick] (-3.5,-5) to [out=-90,in=-90] (3.5,-5);
    \draw[black] plot [mark=*, mark size=3] coordinates{(0,-7.05)};
    
    \draw[thick] (-4.5,-6) -- (-4.5,-5);
    \node[below] at (-4.5,-6) {$A'$};
    \draw[thick] (4.5,-6) -- (4.5,-5);
    \node[below] at (4.5,-6) {$B'$};
    
    \draw[thick] (4.5,5) -- (4.5,6);
    \node[above] at (-4.5,6) {$C$};
    \draw[thick] (-4.5,5) -- (-4.5,6);
    \node[above] at (4.5,6) {$D$};
    
    \end{tikzpicture}
    \caption{}
    \label{fig:non-localcomputation}
    \end{subfigure}
    \caption{Local and non-local computations. a) A channel $\mathcal{T}_{A'B'\rightarrow CD}$ is implemented by directly interacting the input systems. b) A non-local quantum computation. The goal is for the action of this circuit on the $AB$ systems to approximate the channel $\mathcal{T}_{A'B'\rightarrow CD}$.}
    \label{fig:non-localandlocal}
\end{figure*}

In a non-local quantum computation (NLQC), shared entanglement along with a single simultaneous round of quantum communication is used to implement a joint channel. 
See figure \ref{fig:non-localandlocal}. 
NLQC has many applications and connections to other subjects, including in quantum position verification \cite{kent2011quantum,buhrman2014position}, classical information theoretic cryptography \cite{allerstorfer2023relating,asadi2025rank, asadi2025conditional}, communication complexity \cite{girish2026magic}, quantum complexity \cite{speelman2015instantaneous, may2022complexity}, classical complexity \cite{buhrman2013garden,cree2023code}, quantum gravity \cite{may2019quantum,may2020holographic,may2024non,may2022complexity,may2022connected}, uncloneable cryptography \cite{ananth2025unclonable}, Hamiltonian complexity \cite{apel2024security}, and other topics; see \cite{may2026entanglement} for a review. 
We are especially interested in lower bounding the entanglement needed to implement channels in the form of an NLQC. 
In the context of quantum position-verification, entanglement lower bounds on NLQC give guarantee's on how difficult it is to attack a position-verification scheme. 
In the context of quantum gravity, entanglement lower bounds relate to how large of a spacetime region is needed to implement the given quantum channel. 
In fact, in all of the applications of NLQC, understanding the amount of entanglement needed is a key question.

In this article, we are interested in lower bounds on entanglement in NLQC. 
In previous work, most lower bounds have been constructed for specific channels or families of channels \cite{buhrman2014position,tomamichel2013monogamy,junge2021geometry, moreno2026exponential, gonzales2019bounds, asadi2025rank, beigi2011simplified}. 
Here, we are interested in general purpose lower bound techniques. 
Given a channel, we would like to be able to perform a straightforward calculation that lower bounds its entanglement cost. 
Ideally, the lower bound computed would have a simple meaning, and consequently reveal a property of the channel related to entanglement cost. 

We provide a new lower bound technique that relates the entanglement cost of unitaries to the physics of \emph{scrambling}. 
A quantum channel is said to scramble quantum information if information which is initially localized in the inputs becomes delocalized in the outputs. 
Scrambling is closely related to chaos, and has been well-studied in the context of gravity and quantum many-body systems \cite{PRXQuantum.5.010201}.
We find that those unitaries with zero entanglement cost are exactly those which do not scramble information. 
Specifically, the tripartite mutual information was proposed as a measure of scrambling \cite{hosur2016chaos}; we find the tripartite mutual information is zero if and only if the unitary has zero entanglement cost, and further provides a quantitative lower bound on the entanglement cost of a unitary.\footnote{In fact, our bound extends also to general quantum channels, but the interpretation in terms of scrambling is unclear in that case, and the bound is not faithful. For this reason we are primarily interested in applying the bound to unitaries.}

\begin{figure}[htbp]
    \centering
    \begin{tikzpicture}[
        x=1cm, y=1cm,
        line width=0.6pt,
        line cap=round,
        line join=round,
        font=\normalsize
    ]
        \draw (-1.25,0) rectangle (1.25,1.15);
        \node at (0,0.575) {$U_{A'B'\rightarrow CD}$};

        \draw (-0.65,1.15) -- (-0.65,2.05)
            node[above] {$C$};
        \draw ( 0.65,1.15) -- ( 0.65,2.05)
            node[above] {$D$};

        \draw (-1.85,0) node[above] {$A$}
            -- (-1.85,-0.75)
            -- (-1.25,-1.25)
            -- (-0.65,-0.75)
            -- (-0.65,0);
        \node[anchor=west] at (-0.60,-0.90) {$A'$};

        \draw (1.85,0) node[above] {$B$}
            -- (1.85,-0.75)
            -- (1.25,-1.25)
            -- (0.65,-0.75)
            -- (0.65,0);
        \node[anchor=east] at (0.60,-0.90) {$B'$};
    \end{tikzpicture}
    \caption{Choi state of the bipartite unitary
        $U_{A'B'\rightarrow CD}$.
        The lower cups represent maximally entangled states on $AA'$ and $BB'$, and $A,B$ are reference systems.}
    \label{fig:choi-unitary}
\end{figure}
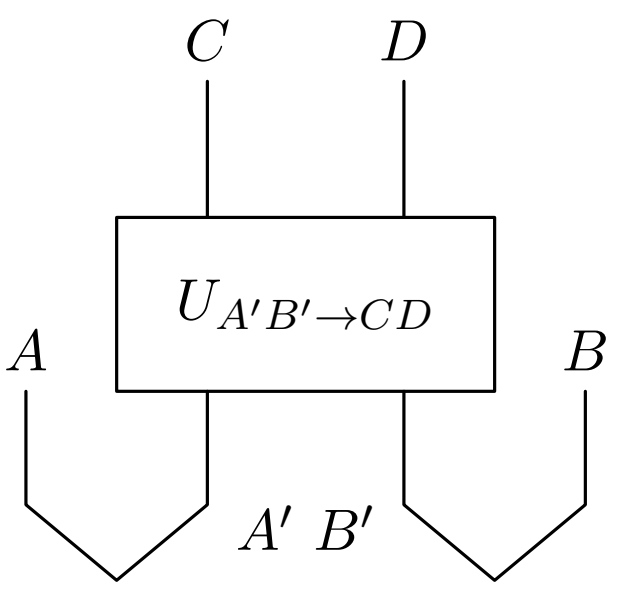

The tripartite mutual information scrambling quantity \cite{hosur2016chaos} is defined as follows. 
Consider a bipartite unitary $U_{A'B'\rightarrow CD}$. 
Define the Choi state of this unitary (see figure \ref{fig:choi-unitary}), 
\begin{align}
    \ket{\chi^U}_{ABCD}=U_{A'B'\rightarrow CD}\ket{\Psi^+}_{AA'}\ket{\Psi^+}_{BB'}
\end{align}
and consider the entropic quantity,
\begin{align}
    J(U)=-I_3(A:C:D)_\chi
\end{align}
where 
\begin{align}\label{eq:I3definition}
    I_3(A:C:D)_\chi:=I(A:C)_\chi+I(A:D)_\chi-I(A:CD)_\chi
\end{align}
and $I(X:Y)$ is the quantum mutual information. 
The quantity $I_3$ is known as the tripartite mutual information. 
We call $J(U)$ the \emph{scrambling quantity}. 
Intuitively, $I_3$ becomes negative, and hence $J(U)$ becomes positive, when the correlation with $A$ is held jointly in $CD$, but not in $C$ or $D$ alone --- in other words, $J(U)$ is positive when the correlation with $A$ has scrambled into $C$ and $D$.
Define $E(U)$ as the minimal entanglement needed to implement $U$ as an NLQC, as quantified by the entanglement of formation in the resource state.
We show that
\begin{align}
    E(U)\geq \frac{J(U)}{2}.
\end{align}
This relates entanglement cost in NLQC and scrambling: Unitaries that are highly scrambling have large entanglement cost in NLQC.
As desired, this bound is straightforward to compute and relates to a simple and physically meaningful property of the unitary.

Leveraging insights from the study of scrambling, we gain an understanding of when this lower bound returns zero, $J(U)=0$. 
As shown in \cite{ding2016conditional}, this occurs exactly when
\begin{align}
    U_{A'B'\rightarrow CD}=(S_{ab\rightarrow C}\otimes T_{a'b'\rightarrow D})(M_{A'\rightarrow aa'}\otimes N_{B'\rightarrow bb'})
\end{align}
for unitaries $S,T,M,N$.
We call such unitaries \emph{subsystem SWAP unitaries}. 
Remarkably, subsystem SWAP unitaries are also implementable with zero entanglement cost: indeed the above explicitly defines an NLQC protocol of the form shown in figure \ref{fig:non-localcomputation}, with no shared entanglement used. 
This means that the scrambling lower bound is faithful: if $J(U)=0$ we have that $U$ is a subsystem SWAP, and hence implementable with zero entanglement cost. 
This is the first known faithful lower bound on entanglement cost for unitaries in NLQC.\footnote{It is open to find faithful lower bounds for quantum channels.}

An earlier lower bound on the entanglement cost of unitaries in NLQC is based on the \emph{controllable correlation}. 
The controllable correlation is defined as follows. 
Consider a unitary $U_{A'B'\rightarrow CD}$ and choose any state $P_{AA'}$, and any state $\phi_{B'}$. 
Then, we consider the state
\begin{align}
    \omega_{AC}(\phi)=\tr_D(U_{A'B'\rightarrow CD}(P_{AA'}\otimes \phi_{B'})U^\dagger_{A'B'\rightarrow CD}).
\end{align}
The controllable correlation is defined by
\begin{align}
    CC^{\opt}(U)=\max_{P,\phi^1,\phi^2}\left(I(A:C)_{\omega(\phi^1)} - I(A:C)_{\omega(\phi^2)}\right).
\end{align}
The work \cite{cleve2026lower} proves that
\begin{align}
    E(U)\geq \frac{CC^{\opt}(U)}{2}.
\end{align}
The controllable correlation reveals another mechanism leading to the unitary $U$ requiring entanglement: If the choice of input on $B'$ can adjust the correlation in $A:C$, then $U$ requires entanglement to implement. 
Further, how much the correlation across $A:C$ can be adjusted provides a quantitative lower bound on entanglement cost. 

A conjecture in \cite{cleve2026lower} asked if the controllable correlation is a faithful measure for entanglement cost. 
Leveraging our results on scrambling lower bounds for entanglement cost, we can prove the controllable correlation is indeed faithful. 
We find that, conveniently, 
\begin{align}
    CC^{\opt}(U)=0\Rightarrow J(U)=0
\end{align}
which suffices to show the controllable correlation is also faithful: when $CC^{\opt}(U)=0$ we have $J(U)=0$, and hence $U$ is a subsystem SWAP, so $U$ has zero entanglement cost.

It is interesting to compare the two lower bound techniques discussed here: the controllable correlation and the scrambling quantity. 
While they are both faithful, and hence both zero for the same set of unitaries, they are nonetheless distinct bounds. 
We give an explicit example where the controllable correlation is larger than the scrambling quantity, and another example where the scrambling quantity is larger than the controllable correlation. 
This reinforces that the two quantities capture qualitatively distinct properties of unitaries that lead to entanglement cost.
Taken together, our two bounds begin to give a general understanding of which properties of a unitary are tied to its entanglement cost as an NLQC. 

\vspace{0.2cm}
\noindent \textbf{Organization of this paper:} In section \ref{sec:QItools} we fix our notation and give some standard quantum information results we make use of in our work.
Section \ref{sec:scramblinglowerboundsection} deals with our lower bound based on scrambling: section \ref{sec:scramblingbackground} gives background on the physics of scrambling, and section \ref{sec:scramblinglowerboundproof} gives the proof of the scrambling based lower bound. 
Section \ref{sec:scramblingboundproperties} then covers a few properties of the bound, including achieving the maximal value of the bound. 
In section \ref{sec:controllablecorrelationbound}, we revisit the controllable correlation lower bound, in particular proving it is faithful.
In section \ref{sec:relationship_CC_scrambling}, we discuss a relationship between the controllable correlation and the scrambling bound, and in particular show they are incomparable. 
We conclude with some open problems and discussion in section \ref{sec:discussion}. 

\section{Background and tools}\label{sec:QItools}

We use $\log$ to denote the base 2 logarithm. 
Hilbert spaces are labelled with capital letters, $\mathcal{H}_A, \mathcal{H}_B,...$, etc. 
Density matrices are labelled with greek letters, $\rho, \sigma, \omega, ...$, etc. 
We use subscripts to track the Hilbert spaces where states live and channels act on, for instance $\mathcal{N}_{A\rightarrow B}(\sigma_A)=\omega_B$.
We use $d_A$ to denote the dimension of Hilbert space $\mathcal{H}_A$, and $n_A$ to denote the log-dimension, $n_A=\log d_A$. 

\vspace{0.2cm}
\noindent \textbf{Entropy:} We define the von Neumann entropy of a density matrix $\rho_A$ as
\begin{align}
    S(A)_\rho = -\tr(\rho_A \log \rho_A).
\end{align}
We define the conditional quantum entropy according to
\begin{align}
    S(A|B)_\rho = S(AB)_\rho - S(B)_\rho.
\end{align}
The conditional entropy is concave in its input state. 
That is for $\rho = \sum_i p_i \rho^i$, we have
\begin{align}
    S(A|B)_\rho \geq \sum_i p_i S(A|B)_{\rho^i}. 
\end{align}
A useful property of the entropy is that for pure states $\psi_{AB}$, the entropy of subsystems is equal, 
\begin{align}
    S(A)_\psi = S(B)_\psi.
\end{align}
When we use this, we say that we exchange subsystems in the entropy using purity. 

\vspace{0.2cm}
\noindent \textbf{Mutual information:} We define the mutual information as
\begin{align}
    I(Q:A)_\rho= S(Q)_\rho+S(A)_\rho-S(QA)_\rho
\end{align}
and the conditional mutual information by
\begin{align}
    I(A:C|B)_\rho=S(AB)_\rho + S(BC)_\rho-S(B)_\rho-S(ABC)_\rho.
\end{align}
The chain rule relates the conditional mutual information to the mutual information, 
\begin{align}
    I(A:C|B)_\rho = I(A:BC)_\rho-I(A:B)_\rho. 
\end{align}
The strong subadditivity inequality expresses that the conditional mutual information is non-negative, $I(A:C|B)_\rho\geq 0$. 
Moving systems into the conditioning argument does not increase the mutual information, 
\begin{align}\label{eq:movetocondition}
    I(A:BC|D)_\rho\geq I(A:B|CD)_\rho.
\end{align}

The conditional mutual information of a classical quantum state satisfies the following. 
\begin{lemma}\label{lem:averagedCMI}
    Consider the quantum state $\rho_{ABX}=\sum_i p_i \rho^i_{AB}\otimes \ketbra{i}{i}_X$. 
    Then
    \begin{align}
        I(A:B|X)_\rho=\sum_i p_i I(A:B)_{\rho^i}
    \end{align}
\end{lemma}

When $I(A:C|B)_\rho=0$, a strong constraint is imposed on the structure of $\rho$. 
\begin{theorem}[\cite{hayden2004structure}]\label{thm:equalityinSSA}
Let $\omega_{ABC}$ be a quantum state. The following are equivalent:
\begin{itemize}
    \item[(1)] $I(A:C|B)_\omega=0$.
    \item[(2)] There exists a decomposition of the $B$ Hilbert space,
    \begin{align}
        \mathcal{H}_B\simeq \bigoplus_i(\mathcal{H}_{L_i}\otimes \mathcal{H}_{R_i})
    \end{align}
    such that $\omega$ can be written in the form
    \begin{align}\label{eq:state_decomposition}
        \omega_{ABC} = \bigoplus_i p_i \,\omega^i_{AL_i} \otimes \omega^i_{R_iC}.
    \end{align}
    \item[(3)] There exists a channel $\mathcal{R}_{B \rightarrow AB}$ such that $\omega_{ABC}= (\mathcal{R}_{B \rightarrow AB} \otimes \id_C) (\omega_{BC})$.
\end{itemize}
\end{theorem}

We will use the data processing inequality for mutual information, which states that the mutual information $I(A:B)$ cannot increase under any local channel acting on $A$ and $B$.

\begin{lemma}\label{lem:dataprocessinginequality}
    Let $\rho_{AB}$ be a quantum state, and let $\mathcal N_{A\to A'}$ and $\mathcal M_{B\to B'}$ be quantum channels. Let $\sigma_{A'B'C} =(\mathcal N_{A\to A'} \otimes \mathcal M_{B\to B'})(\rho_{ABC})$. Then,
    \begin{align}
        I(A:B|C)_\rho \geq I(A':B'|C)_\sigma.
    \end{align}
\end{lemma}

\vspace{0.2cm}
\noindent \textbf{Quantum channels:} Given a quantum channel $\mathcal{N}_{B\rightarrow C}$, we will make use of the Choi state of the channel, 
\begin{align}
    \chi_{AC}^{\mathcal{N}} = \mathcal{N}_{B\rightarrow C}(\Psi^+_{AB}) = \frac{1}{d_A} \sum_{i,j}\ketbra{i}{j}_A\otimes \mathcal{N}_{B\rightarrow C}(\ketbra{i}{j}_B)
\end{align}
We note that the Choi state of a channel uniquely fixes the channel.

We also use the diamond norm distance on channels. 
\begin{definition} Let $\mathcal{N}_{B\rightarrow C}, \mathcal{M}_{B\rightarrow C}: \mathcal{L}(\mathcal{H}_B)\rightarrow \mathcal{L}(\mathcal{H}_C)$ be quantum channels. 
The \textbf{diamond norm distance} is defined by 
\begin{align}
    \Vert\mathcal{N}_{B\rightarrow C}-\mathcal{M}_{B\rightarrow C}\Vert_\diamond = \sup_{d} \max_{\Psi_{A_dB}}\Vert\mathcal{N}_{B\rightarrow C}(\Psi_{A_dB}) - \mathcal{M}_{B\rightarrow C}(\Psi_{A_dB})\Vert_1
\end{align}
where $\mathcal{H}_{A_d}$ is a $d$ dimensional Hilbert space. 
\end{definition}

\vspace{0.2cm}
\noindent \textbf{Entanglement:} We make use of the following measure of entanglement. 
\begin{definition}
    The \textbf{entanglement of formation} is defined as
\begin{align}
    E_f(A:B)_\rho = \min_{\{p_i,\ket{\psi_i}\}} \sum_i p_i S(B)_{\psi_i},
\end{align}
where the minimization is over ensembles $\{p_i,\ket{\psi_i}\}$ such that $\rho=\sum_i p_i\ketbra{\psi_i}{\psi_i}$. 
\end{definition}
Note that this is a faithful measure: $E_f(A:B)_\rho=0$ if and only if $\rho$ is separable. 

\section{A faithful lower bound from scrambling}\label{sec:scramblinglowerboundsection}

\subsection{Scrambling, tripartite mutual information, and zero-entanglement unitaries}\label{sec:scramblingbackground}

To develop our lower bound strategy our starting point is the observation that at least the following class of unitaries, mentioned in the introduction, can be implemented as NLQCs using zero entanglement. 

\begin{definition}
    A unitary $U_{A'B'\rightarrow CD}$ is said to be a \textbf{subsystem SWAP} unitary if it takes the form
    \begin{align}\label{eq:subsystemSWAPunitary}
        U_{A'B'\rightarrow CD}=(S_{ab\rightarrow C}\otimes T_{a'b'\rightarrow D})(M_{A'\rightarrow aa'}\otimes N_{B'\rightarrow bb'}).
    \end{align}
    where $S,T,M,N$ are unitaries.
\end{definition}
From the form of the NLQC circuit, it is apparent any subsystem SWAP unitary can be implemented without using any entanglement. 
It is intuitive, but not entirely apparent, that every unitary that can be implemented with zero entanglement should be a subsystem SWAP unitary; plausibly the use of local channels rather than local unitaries could allow a larger class of unitaries to be implemented. 
Since the overall action is unitary however, we might guess that the subsystem SWAP form accounts for all zero-entanglement cost unitaries. 
Motivated by this, we begin investigating how to characterize the subsystem SWAP unitaries. 

The subsystem SWAP unitaries appear in the scrambling literature, in the context of the following theorem from \cite{ding2016conditional}. 

\begin{theorem}\label{thm:DingHaydenWalter}
    Let $U_{A'B'\rightarrow CD}$ be a unitary, and define the Choi state for this unitary 
    \begin{align}
        \ket{\chi^U}_{ABCD}=U_{A'B'\rightarrow CD}\ket{\Psi^+}_{A'A}\ket{\Psi^+}_{B'B}.
    \end{align} 
    Then, 
    \begin{align}
        I(A:B|C)_{\chi^U}=0 \,\Leftrightarrow \, U_{A'B'\rightarrow CD} \,\, \text{is a subsystem SWAP unitary.}
    \end{align}
\end{theorem}
In the Choi state $\ket{\chi^U}$, because the density matrix on $AB$ is maximally mixed and the state on $ABCD$ is pure, it follows from the definition of the tripartite mutual information (equation \eqref{eq:I3definition}) that
\begin{align}
    I_3(A:C:D)_{\chi^U} &= -I(A:B|C)_{\chi^U}.
\end{align}
Thus we can also view the subsystem SWAP unitaries as those whose Choi state has zero tripartite mutual information. 
This motivates us to recall the previous reasoning for studying the tripartite mutual information, which relates to the physics of scrambling.

From the definition of the tripartite mutual information, we see that it is negative when
\begin{align}
    I(A:CD)>I(A:C)+I(A:D).
\end{align}
This is saying that it is negative when more correlation with $A$ is stored jointly in $CD$ than in $C$ and $D$ taken separately. 
Considering the subsystem SWAP unitaries, it is clear they do not scramble information in this sense: ignoring the local unitaries (which do not change the correlation measures), the subsystem SWAP keeps information localized, and just rearranges where certain qubits sit. 

It is also interesting to recall the relationship between scrambling and chaos found in \cite{hosur2016chaos}, which is phrased in terms of the tripartite mutual information.
To understand the relationship between chaos and scrambling, consider a complete basis of operators on the $A'$ Hilbert space, call them $\{\mathcal{O}^i_{A'}\}_i$, and a complete basis of operators on the $D$ Hilbert space, call them $\{\mathcal{O}^i_{D}\}_i$. 
A standard diagnostic of chaos in the unitary $U_{A'B'\rightarrow CD}$ is the \emph{out-of-time-order correlation function}, or OTOC, defined as
\begin{align}
    \langle \mathcal{O}^i(t)\mathcal{O}^j_D\mathcal{O}^i(t)\mathcal{O}^j_D\rangle =\tr(\rho_{CD}\mathcal{O}^i(t)\mathcal{O}^j_D\mathcal{O}^i(t)\mathcal{O}^j_D).
\end{align}
Here, by $\mathcal{O}^i(t)$ we mean the time evolved operator, $\mathcal{O}^i(t):=U_{A'B'\rightarrow CD}\mathcal{O}^i_{A'} U^\dagger_{A'B'\rightarrow CD}$, and by $\rho_{CD}$ we mean the $CD$ subsystem of the Choi state.\footnote{This is just the maximally mixed state so $\rho_{CD}$ just adds a normalization factor here, though in more general contexts this is replaced with a finite temperature thermal state.} 
We are specifically interested in the value of this correlation function averaged over choice of operator,
\begin{align}
    \langle \mathcal{O}^i(t)\mathcal{O}^j_D\mathcal{O}^i(t)\mathcal{O}^j_D\rangle_{avg}=\frac{1}{d_{A'}^2d_D^2}\sum_{i,j}\langle \mathcal{O}^i(t)\mathcal{O}^j_D\mathcal{O}^i(t)\mathcal{O}^j_D\rangle.
\end{align}
This averaged OTOC is related in \cite{hosur2016chaos} to the tripartite mutual information according to
\begin{align}
    c_02^{I_3(A:C:D)_{\chi^U}/2} \leq \langle \mathcal{O}^i(t)\mathcal{O}^j_D\mathcal{O}^i(t)\mathcal{O}^j_D\rangle_{avg}.
\end{align}
Here $c_0$ is a constant related to the minimal values of $I_3$ and the averaged correlation function; we omit these details for simplicity. 
We see that if the OTOC becomes small, a signature that the evolution of $U$ is chaotic, then $I_3$ must become negative. 

Another appearance of the tripartite mutual information is in holography \cite{hayden2013holographic}, where it is found that holographic states have $I_3\leq 0$ always. 
Intriguingly, NLQC has also been related to holography in \cite{may2019quantum, may2020holographic, may2022complexity}, although its role there is not obviously related to the tripartite mutual information or scrambling.

\subsection{Lower bound on entanglement from scrambling}\label{sec:scramblinglowerboundproof}

We are now ready to prove that scrambling gives a lower bound on the entanglement cost of implementing a unitary as an NLQC. 
We first give a formal definition of the scrambling quantity. 

\begin{definition}\label{def:scramblingvalue}
    We define the \textbf{scrambling value} $J(\mathcal{N})$ of a quantum channel $\mathcal{N}_{A'B'\rightarrow CD}$ as follows. 
    Take the Choi state of $\mathcal{N}_{A'B'\rightarrow CD}$, labelled $\chi^{\mathcal{N}}_{ABCD}=\mathcal{N}_{A'B'\rightarrow CD}({\Psi^+}_{AA'}\otimes {\Psi^+}_{BB'})$, and then define
    \begin{align}
        J(\mathcal{N}):= S(BC)_{\chi^{\mathcal{N}}}+S(BD)_{\chi^{\mathcal{N}}}-S(ABC)_{\chi^{\mathcal{N}}}-S(ABD)_{\chi^{\mathcal{N}}}. 
    \end{align}
\end{definition}
This quantity is defined so as to appear as the lower bound in our proof below; we call it the scrambling quantity however because of its connection to the tripartite mutual information and scrambling when $\mathcal{N}$ is taken to be a unitary. 
Specifically, if we choose a unitary channel so that the Choi state is pure, we have
\begin{align}
    J(U)&=S(BC)_{\chi^{U}}+S(BD)_{\chi^{U}}-S(ABC)_{\chi^{U}}-S(ABD)_{\chi^{U}} \nonumber \\
    &=S(BC)_{\chi^{U}}+S(AC)_{\chi^{U}}-S(ABC)_{\chi^{U}}-S(C)_{\chi^{U}} \nonumber \\
    &=I(A:B|C)_{\chi^U} \nonumber \\
    &=-I_3(A:C:D)_{\chi^U}.
\end{align}
Thus specialized to unitaries our ``scrambling quantity'' is exactly the tripartite mutual information relevant to scrambling. 

Before proving our lower bound, we need the following property of the scrambling quantity. 
\begin{lemma}\label{lem:Jconvex}
    The scrambling quantity is convex, that is
    \begin{align}
        J\left(\sum_i p_i \mathcal{N}^i\right) \leq \sum_i p_i J\left( \mathcal{N}^i\right).
    \end{align}
\end{lemma}
\begin{proof}\,
    Let $\mathcal{N}=\sum_i p_i \mathcal{N}^i$. 
    Then notice that
    \begin{align}
        \chi^{\mathcal{N}}=\sum_i p_i \chi^{\mathcal{N}^i}
    \end{align}
    Now re-express the scrambling quantity as
    \begin{align}
        J(\mathcal{N}) = -S(A|BC)_{\chi^\mathcal{N}}-S(A|BD)_{\chi^\mathcal{N}}
    \end{align}
    Now using that the conditional entropy is concave in the input state, we have that $J$ is convex, as claimed. 
\end{proof}

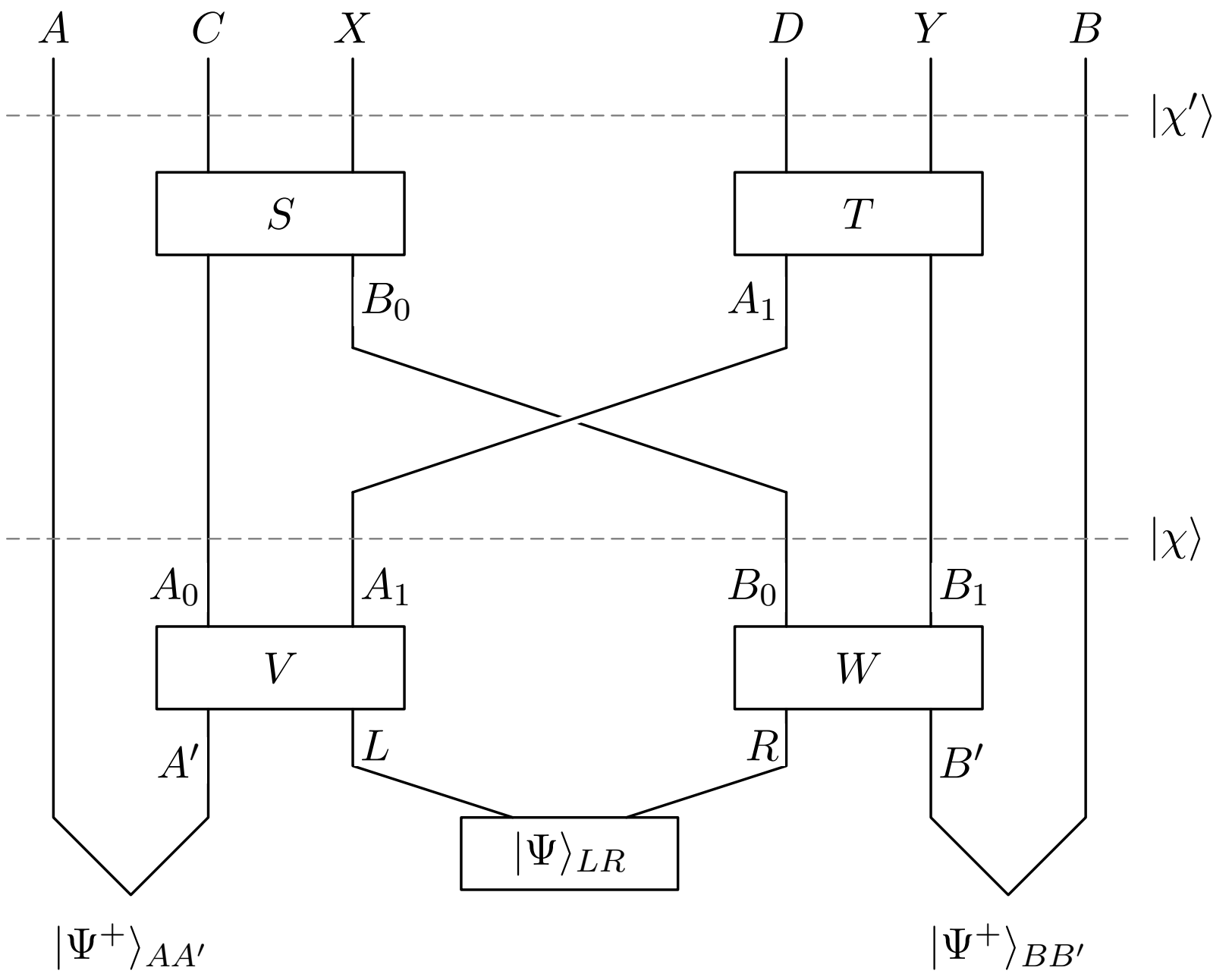
\begin{figure}[htbp]
    \centering
    \begin{tikzpicture}[
        x=0.9cm, y=0.9cm,
        line width=0.65pt,
        line cap=round,
        line join=round,
        font=\normalsize,
        reg/.style={fill=white,inner sep=1.5pt}
    ]
        \draw (-4.0,1.8) rectangle (-1.6,2.6);
        \node at (-2.8,2.2) {$V$};
        \draw (1.6,1.8) rectangle (4.0,2.6);
        \node at (2.8,2.2) {$W$};
        \draw (-4.0,6.2) rectangle (-1.6,7.0);
        \node at (-2.8,6.6) {$S$};
        \draw (1.6,6.2) rectangle (4.0,7.0);
        \node at (2.8,6.6) {$T$};

        \draw (-5.0,8.1) node[above] {$A$}
            -- (-5.0,0.75) -- (-4.25,0)
            -- (-3.5,0.75) -- (-3.5,1.8);
        \node[reg,left] at (-3.5,1.3) {$A'$};
        \node[below=3pt] at (-4.25,0) {$|\Psi^+\rangle_{AA'}$};

        \draw (5.0,8.1) node[above] {$B$}
            -- (5.0,0.75) -- (4.25,0)
            -- (3.5,0.75) -- (3.5,1.8);
        \node[reg,right] at (3.5,1.3) {$B'$};
        \node[below=3pt] at (4.25,0) {$|\Psi^+\rangle_{BB'}$};

        \draw (-1.05,0.05) rectangle (1.05,0.75);
        \node at (0,0.4) {$|\Psi\rangle_{LR}$};
        \draw (-0.55,0.75) -- (-2.1,1.25) -- (-2.1,1.8);
        \draw ( 0.55,0.75) -- ( 2.1,1.25) -- ( 2.1,1.8);
        \node[reg,right] at (-2.1,1.45) {$L$};
        \node[reg,left]  at ( 2.1,1.45) {$R$};

        \draw (-3.5,2.6) -- (-3.5,6.2);
        \draw ( 3.5,2.6) -- ( 3.5,6.2);

        \draw (-2.1,2.6) -- (-2.1,3.9);
        \draw ( 2.1,2.6) -- ( 2.1,3.9);
        \node[reg,left]  at (-3.5,2.98) {$A_0$};
        \node[reg,right] at (-2.1,2.98) {$A_1$};
        \node[reg,left]  at ( 2.1,2.98) {$B_0$};
        \node[reg,right] at ( 3.5,2.98) {$B_1$};

        \draw (2.1,3.9) -- (-2.1,5.3);
        \draw[white,line width=3pt] (-0.3,4.5) -- (0.3,4.7);
        \draw (-2.1,3.9) -- (2.1,5.3);
        \draw (-2.1,5.3) -- (-2.1,6.2);
        \draw ( 2.1,5.3) -- ( 2.1,6.2);
        \node[reg,right] at (-2.1,5.75) {$B_0$};
        \node[reg,left]  at ( 2.1,5.75) {$A_1$};

        \draw[densely dashed,gray,line width=0.45pt]
            (-5.45,3.45) -- (5.45,3.45);
        \node[right] at (5.45,3.45) {$|\chi\rangle$};

        \draw (-3.5,7.0) -- (-3.5,8.1) node[above] {$C$};
        \draw (-2.1,7.0) -- (-2.1,8.1) node[above] {$X$};
        \draw ( 2.1,7.0) -- ( 2.1,8.1) node[above] {$D$};
        \draw ( 3.5,7.0) -- ( 3.5,8.1) node[above] {$Y$};

        \draw[densely dashed,gray,line width=0.45pt]
            (-5.45,7.55) -- (5.45,7.55);
        \node[right] at (5.45,7.55) {$|\chi'\rangle$};

    \end{tikzpicture}
    \caption{An NLQC protocol with pure shared resource $|\Psi\rangle_{LR}$. The reference-input pairs $AA'$ and $BB'$ are maximally entangled. The dashed lines mark $\ket{\chi}_{AA_0A_1BB_0B_1}$ and $\ket{\chi'}_{ABCDXY}$.
    The systems $A_1$ and $B_0$ are exchanged before $S$ and $T$ produce $C,D,X,Y$. 
    The reference systems $A,B$ remain untouched.}
    \label{fig:nlqc-protocol}
\end{figure}

When we lower bound the entanglement cost of implementing an NLQC for a given channel, we mean entanglement cost in the following sense. 
\begin{definition}\label{def:entanglementcost}
    Consider a channel $\mathcal{N}_{A'B'\rightarrow CD}$. Then we define the cost to implement $\mathcal{N}_{A'B'\rightarrow CD}$ as an NLQC to be the minimal value of $E_f(L:R)_\Psi$, minimized over all resource states $\Psi_{LR}$ that suffice to implement $\mathcal{N}$ exactly. 
\end{definition}

Finally, we are ready to prove the lower bound on entanglement cost based on the scrambling quantity. 

\begin{theorem}\label{thm:Jlowerbound}
For any channel $\mathcal{N}$, let $E(\mathcal{N})$ denote the entanglement cost of $\mathcal{N}$ (definition \ref{def:entanglementcost}) and $J(\mathcal{N})$ the scrambling value (definition \ref{def:scramblingvalue}). 
Then
\begin{align}
    E(\mathcal{N}) \geq \frac{1}{2}J(\mathcal{N}).
\end{align}
\end{theorem}
\begin{proof}\,
Recall that
\begin{align}
    E(\mathcal{N})=\min_{\Psi} E_f(L:R)_{\Psi}
\end{align}
where the minimization is over all states $\Psi_{LR}$ that can be used as resource states to implement $\mathcal{N}_{A'B'\rightarrow CD}$ as an NLQC. 

We first consider a pure resource state $\Psi_{LR}$ and derive a bound on the entropy of $L$.
Without loss of generality, we assume the local operations in the NLQC protocol are isometric.
Let the local operations and intermediate subsystems be labelled according to:
\begin{align}
    V_{A'L\rightarrow A_0A_1}, W_{B'R\rightarrow B_0B_1}, S_{A_0B_0\rightarrow CX}, T_{A_1B_1\rightarrow DY}.
\end{align}
These operations and the labels are shown in figure \ref{fig:nlqc-protocol}. 
Consider the state
\begin{align}
    \ket{\chi}_{AA_0A_1BB_0B_1}=(V_{A'L\rightarrow A_0A_1}\otimes W_{B'R\rightarrow B_0B_1})\ket{\Psi^+}_{AA'}\ket{\Psi^+}_{BB'}\ket{\Psi}_{LR}.
\end{align}
First notice that, since local isometries cannot change the mutual information, 
\begin{align}\label{eq:2SMI}
    2S(L)_\Psi=I(L:R)_\Psi=I(AA_0A_1:BB_0B_1)_\chi.
\end{align}
Now, use that
\begin{align}
    I(AA_0A_1:BB_0B_1)_\chi\geq I(AA_0:BB_0)_\chi \geq I(A:B|A_0B_0)_\chi = I(A:B|CX)_{\chi'}
\end{align}
where we define the final state as
\begin{align}
    \ket{\chi'}_{ABCDXY} :=
    \left(
        I_{AB}\otimes S_{A_0B_0\rightarrow CX}\otimes T_{A_1B_1\rightarrow DY}
    \right)
    \ket{\chi}_{AA_0A_1BB_0B_1}.
\end{align}
The first inequality uses data processing (lemma \ref{lem:dataprocessinginequality}) applied to the partial trace, the second inequality uses equation \eqref{eq:movetocondition} twice, and the last equality uses that the map from $A_0B_0$ to $CX$ is isometric.
Combining this with equation \eqref{eq:2SMI}, we get
\begin{align}\label{eq:2SIc}
    2S(L)_\Psi \geq I(A:B|CX)_{\chi'}.
\end{align}
Since $X$ is an output to the environment and isn't fixed by the channel $\mathcal{N}$ (but rather by the isometric extension of $\mathcal{N}$), we want to remove it from our bound on $S(L)$, to obtain a bound that depends only on $\mathcal{N}$. 
To do this, we use that
\begin{align}
    I(A:B|CX)_{\chi'} &= I(A:BCX)_{\chi'} - I(A:CX)_{\chi'} \nonumber \\
    &\geq I(A:BC)_{\chi'} - I(A:CX)_{\chi'} 
    \geq I(A:BC)_{\chi'} - I(A:CXY)_{\chi'}.
\end{align}
The two inequalities hold from the data processing inequality (lemma \ref{lem:dataprocessinginequality}).
Then using purity to exchange subsystems and because $A$ is maximally mixed, we get that $I(A:CXY)_{\chi'}=2n_A-I(A:BD)_{\chi'}$, so then
\begin{align}
    I(A:B|CX)_{\chi'} \geq I(A:BC)_{\chi'} + I(A:BD)_{\chi'} - 2n_A.
\end{align}
Combining this with equation \eqref{eq:2SIc}, we get
\begin{align}\label{eq:entropyinequality_finalpurestate}
    2S(L) \geq I(A:BC)_{\chi'} + I(A:BD)_{\chi'} - 2n_A.
\end{align}
From the definitions
\begin{align}
    \chi^{\mathcal N}_{ABCD} =\tr_{XY} ({\chi'}_{ABCDXY}).
\end{align}
Using that the density matrix on $A$ is maximally mixed, \eqref{eq:entropyinequality_finalpurestate} is equivalently
\begin{align}
    2S(L)_{\Psi} \geq S(BC)_{\chi^{\mathcal{N}}}+S(BD)_{\chi^{\mathcal{N}}}-S(ABC)_{\chi^{\mathcal{N}}}-S(ABD)_{\chi^{\mathcal{N}}} = J(\mathcal{N}). 
\end{align}
Thus we have
\begin{align}\label{eq:purestateJbound}
    {S(L)_{\Psi} \geq \frac{J(\mathcal{N})}{2}}
\end{align}
for any pure resource state $\Psi$. 

Next, we combine this with the definition of the entanglement of formation and convexity of $J$ to obtain a lower bound on entanglement cost. 
Suppose that the resource state is a mixed state $\Psi=\sum_i p_i \Psi^i$, where we consider the ensemble decomposition of $\Psi$ that minimizes $\sum_i p_i S(L)_{\Psi_i}$, i.e. the minimizing ensemble in the definition of the entanglement of formation.  
Then the channel implemented by the NLQC protocol will be $\mathcal{N}=\sum_i p_i \mathcal{N}^i$. 
To implement $\mathcal{N}^i$, from equation \eqref{eq:purestateJbound} we get that we must have
\begin{align}
    S(L)_{\Psi^i}\geq \frac{J(\mathcal{N}^i)}{2}.
\end{align}
 Now we use this to bound the entanglement of formation, 
\begin{align}
    E_f(L:R)_\Psi=\sum_i p_i S(L)_{\Psi^i}\geq \sum_i p_i\frac{J(\mathcal{N}^i)}{2} \geq  \frac{J(\mathcal{N})}{2},
\end{align}
where the last inequality follows from the convexity of the scrambling quantity (lemma \ref{lem:Jconvex}). This concludes the proof.
\end{proof}

In the context of controllable correlation, only unitaries are given lower bounds. 
As a consequence, \cite{cleve2026lower} spends considerable effort in proving that channels which are close to unitary also obtain lower bounds. 
Here, since the bound applies to channels, we can circumvent this analysis: for a channel which is close to a unitary we can obtain a bound directly by evaluating the scrambling quantity for that channel. 
Further, since $J(\mathcal{N})$ is defined in terms of entropy of the Choi state, the entropy is continuous in the input state, and the Choi state is continuous in the channel, we know nearby channels obtain similar lower bounds immediately. 

Focusing on the unitary case we find the following theorem, which leads to a complete characterization of unitary channels with zero entanglement cost. 
\begin{theorem}
    Considering only unitary channels, the lower bound of theorem \ref{thm:Jlowerbound} is faithful; that is, 
    \begin{align}
        J(U)=0 \Leftrightarrow U\, \text{has zero entanglement cost}
    \end{align}
\end{theorem}
\begin{proof}\,
    That $J(U)>0$ implies there is an entanglement cost is given by theorem \ref{thm:Jlowerbound}. 
    That $J(U)=0$ implies the entanglement cost is zero follows from theorem \ref{thm:DingHaydenWalter}: for $U$ unitary, $J(U)=0$ means that $U$ is a subsystem swap, which can be directly implemented as an entanglement free NLQC. 
\end{proof}

Returning to our discussion of scrambling, we see that a unitary $U_{A'B'\rightarrow CD}$ has zero entanglement cost if and only if it does not scramble information into $CD$. 
This gives a simple and physical interpretation of when unitaries have an entanglement cost in NLQC. 
Further, the lower bound of theorem \ref{thm:Jlowerbound} instructs us that unitaries that more thoroughly scramble information have larger entanglement costs.

\subsection{Properties of the scrambling bound}\label{sec:scramblingboundproperties}

\subsubsection*{Bound maximizing unitaries}

We have characterized the cases where, for a unitary $U$, $J(U)=0$. 
This occurs exactly when $U$ is a subsystem SWAP unitary. 
It is also interesting to understand the maximal value of $J(U)$ and understand when this maximum is saturated. 
In fact this follows immediately from a discussion in \cite{ding2016conditional}, who understand when unitaries are maximally scrambling in the case where $d=d_{A'}=d_{B'}=d_C=d_D$. 
For convenience we recall the most relevant points of that discussion briefly here. 

First recall that
\begin{align}
    J(U)&=I(A:B|C)_{\chi^U} 
\end{align}
so that
\begin{align}
    J(U)\leq 2 n_A.
\end{align}
But also, 
\begin{align}
    I(A:B|C)_{\chi^U} &= S(AC)_{\chi^U} +S(BC)_{\chi^U}-S(C)_{\chi^U}-S(ABC)_{\chi^U} \nonumber \\
    &=S(AC)_{\chi^U} +S(AD)_{\chi^U}-S(C)_{\chi^U}-S(D)_{\chi^U} \nonumber \\
    &=2S(A)_{\chi^U}-I(A:C)_{\chi^U} - I(A:D)_{\chi^U} \nonumber \\
    &=2n_A-I(A:C)_{\chi^U} - I(A:D)_{\chi^U}.
\end{align}
This tells us that to achieve maximal scrambling, we need $I(A:C)=I(A:D)=0$.
But now notice that by the definition of $\chi^U$, both $\chi^{U}_{AB}$ and $\chi^U_{CD}$ are maximally mixed, so every single system marginal is maximally mixed. 
But then if $\chi^{U}_C, \chi^{U}_A$ are both maximally mixed and $I(A:C)_{\chi^{U}}=0$, it follows that $\chi^U_{AC}$ is maximally mixed. 
Similarly $\chi^U_{AD}$ is maximally mixed. 
Using $S(BC)_{\chi^U}=S(AD)_{\chi^U}$ and that $\chi^U_{AD}$ is maximally mixed, we get that $BC$ is maximally mixed. 
Similarly $S(BD)_{\chi^U}=S(AC)_{\chi^U}$ and $\chi^U_{AC}$ we already found was maximally mixed, so $BD$ is as well. 
Thus every two subsystem marginal of the Choi state $\chi^{U}_{ABCD}$ is maximally mixed, and hence is also maximally entangled across every bipartition. 

If a state
\begin{align}
    \ket{\Psi}_{ABCD} = \sum_{ijk\ell} c_{ijk\ell} \ket{i}_A\ket{j}_B\ket{k}_C\ket{\ell}_D
\end{align}
is maximally entangled across every bipartition, we say that $c_{ijk\ell}$ is a \emph{perfect tensor}. 
Given this understanding, we can straightforwardly use perfect tensors to construct examples of $J(U)$ maximizing unitaries. 
An example is the unitary, defined for odd $d$, 
\begin{align}\label{eq:perfectunitary}
    U_{A'B'\rightarrow CD}^\text{perfect}\ket{i}_{A'}\ket{j}_{B'}=\ket{i+j}_C\ket{i-j}_D
\end{align}
where all arithmetic is modulo $d$. 
This unitary is defined by a perfect tensor and hence is maximally scrambling, i.e., $J(U^\text{perfect})=2\log d$.
Some further discussion of the information-theoretic properties of maximally scrambling unitaries can be found in \cite{ding2016conditional}.

\subsubsection*{Unitaries with small values of $J(U)$}

While we showed that $J(U)=0$ means the unitary $U$ is a subsystem SWAP unitary, and hence can be implemented with zero entanglement as an NLQC, it is natural to ask if $J(U)$ small implies that $U$ must be close to the subsystem SWAP form. 
This question is actually also already understood in the scrambling literature \cite{ding2016conditional}. 
We recall their insights that address this question here. 

It is possible to construct a sequence of unitaries $\{U_d\}_d$, one for each choice of local dimension $d$, such that as $d$ becomes large $J(U_d)$ goes to zero, even while $U_d$ stays far (in diamond norm) from any subsystem SWAP unitary. 
This is the content of the following theorem, proven in \cite{ding2016conditional}.

\begin{theorem}\label{thm:smallJunitary}
Let $d_S$ be an odd constant. Then there is a sequence of bipartite unitaries $\{U_d\}_d$ such that
\begin{align}
    \lim_{d\rightarrow \infty}J(U_d)=0
\end{align}
and 
\begin{align}
    \liminf_{d\rightarrow \infty} \inf_{U_0:J(U_0)=0} \Vert U_d-U_0\Vert_\diamond \geq 1.
\end{align}
\end{theorem}
The construction of the unitaries $U_d$ is as follows. 
Let $\ket{a}, \ket{b}$ be basis vectors in $d$ dimensional spaces. 
Then define $U_d$ to act according to
\begin{align}
    U_d\ket{a}_{A}\ket{b}_B=
    \begin{cases}
        U_S\ket{a}_{A'}\ket{b}_{B'} \,\,\,\text{  if  } 0\leq a,b \leq d_S-1\nonumber \\
        \ket{a}_{A'}\ket{b}_{B'} \quad \quad \,\,\,\,\,\,\,\,\text{otherwise}
    \end{cases}
\end{align}
where $U_S$ is any maximally scrambling unitary and $d_S$ is a fixed constant.
Thus $U_d$ acts as a maximally scrambling unitary on a subspace whose relative size goes to zero as $d$ becomes large. 

\section{Faithfulness of the controllable correlation}\label{sec:controllablecorrelationbound}

In this section we relate the scrambling bound and the earlier bound based on controllable correlation \cite{cleve2026lower}. 
This allows us to prove that the controllable correlation is a faithful lower bound. 
We begin with recalling the definition of the controllable correlation and investigating some of its properties. 

\subsection{Basic properties of the controllable correlation}

We first recall the definition of the controllable correlation. 
\begin{definition}
    Consider a unitary $U_{A'B'\rightarrow CD}$ and choose a state on $AA'$, which we label $P_{AA'}$, with $n_{A'}=n_A$.
    Define the states
    \begin{align}
        \rho^1_{ACD}&=U_{A'B'\rightarrow CD}(P_{AA'}\otimes \phi^1_{B'})U_{A'B'\rightarrow CD}^\dagger, \nonumber \\
        \rho^2_{ACD}&=U_{A'B'\rightarrow CD}(P_{AA'}\otimes \phi^2_{B'} )U_{A'B'\rightarrow CD}^\dagger.
    \end{align}
    We say that $U_{A'B'\rightarrow CD}$ has $(\lambda_1,\lambda_2)$-controllable correlation if there exist states $\phi^1_{B'}, \phi^2_{B'}$, and $P_{AA'}$ such that
    \begin{align}
        \lambda_1=I(A:C)_{\rho^1}, \qquad\lambda_2=I(A:C)_{\rho^2}.
    \end{align}
    We always consider the case where $\lambda_1\geq \lambda_2$. If there is no choice of states $P_{AA'}$, $\phi_{B'}^1,\phi_{B'}^2$ such that $\lambda_1>\lambda_2$ we say $U_{A'B'\rightarrow CD}$ is not controllably correlated. 
\end{definition}
It will be useful to define the quantity
\begin{align}
    CC(U,P, \phi^1, \phi^2)=I(A:C)_{\rho^1}-I(A:C)_{\rho^2}
\end{align}
and then
\begin{align}
    CC^{\opt}(U) = \max_{P, \phi^1, \phi^2} CC(U,P, \phi^1, \phi^2).
\end{align}

In \cite{cleve2026lower}, the following lower bound is proven. 
\begin{theorem}\label{thm:controllablecorrelationlowerbound}
    If an NLQC protocol using a resource $\Psi$ gives a perfectly correct implementation of $U_{A'B' \rightarrow CD}$, then
    \begin{align}
        E_f(L:R)_\Psi \geq \frac{CC^{\opt}(U)}{2}.
    \end{align}
\end{theorem}
Note that we focus on the perfect case, while \cite{cleve2026lower} allows for an $\epsilon$-correct implementation of $U_{A'B'\rightarrow CD}$ and then proves a lower bound that depends on $\epsilon$. 

We define a simpler quantity than the optimized controllable correlation, which fixes $P_{AA'}=\Psi^+_{AA'}$, the maximally entangled state, and does not involve an optimization.
\begin{definition}
    Consider any ensemble $E=\{(p_i, \phi^i_{B'})\}_{i=1}^{r}$.
    The \textbf{averaged controllable correlation} using ensemble $E$ is defined by
    \begin{align}
        \overline{CC}(U,E)=\sum_i p_i I(A:C)_{\rho^i}-I(A:C)_{\rho^0}
    \end{align}
    where, for $i\geq 1$, 
    \begin{align}
        \rho^i_{ACD}=U_{A'B'\rightarrow CD}(\Psi^+_{AA'}\otimes \phi^i_{B'}) U_{A'B'\rightarrow CD}^\dagger. 
    \end{align}
    and by $\rho^0$ we mean $U_{A'B'\rightarrow CD}(\Psi^+_{AA'}\otimes \pi_{B'}) U_{A'B'\rightarrow CD}^\dagger$. 
\end{definition}

The averaged controllable correlation can also be viewed as a conditional mutual information, as long as the ensemble satisfies $\sum_i p_i \phi^i_{B'}=\pi_{B'}$, where $\pi_B$ is the maximally mixed state.
\begin{lemma}\label{lem:averageCC_CMI}
    Consider any ensemble $E=\{(p_i, \phi^i_{B'})\}$ with $\sum_i p_i \phi^i_{B'}=\pi_{B'}$, where $\pi_{B'}$ is the maximally mixed state.
    Define
    \begin{align}
        \omega_{ACX} = \sum_i p_i \rho^i_{AC}\otimes \ketbra{i}{i}_X
    \end{align}
    where $\rho^i_{AC}=\tr_{D}(U_{A'B'\rightarrow CD}(\Psi^+_{AA'}\otimes \phi^i_{B'})U^\dagger_{A'B'\rightarrow CD})$, so that $\omega_{ACX}$ is the output of the unitary if the $B'$ subsystem of the state $\sum_ip_i\phi_{B'}^i\otimes \ketbra{i}{i}_X$ is taken as the input on $B'$. 
    Then we have
    \begin{align}
        \overline{CC}(U,E) = I(A:X|C)_\omega
    \end{align}
\end{lemma}
\begin{proof}\,
    By the chain rule, we have
    \begin{align}\label{eq:chainruleQAX}
        I(A:X|C)_\omega = I(A:CX)_\omega-I(A:C)_\omega.
    \end{align}
    Notice that 
    \begin{align} 
        \omega_{AC} &=\tr_{X} \left( \sum_i p_i \rho^i_{AC}\otimes \ketbra{i}{i}_X\right)\nonumber \\
        &=\sum_i p_i \rho^i_{AC} \nonumber \\
        &= \sum_i p_i \tr_D\left( U_{A'B' \rightarrow CD}(\Psi^+_{AA'}\otimes \phi^i_{B'})U^\dagger_{A'B'\rightarrow CD}\right) \nonumber \\
        &= \tr_D\left( U_{A'B' \rightarrow CD}(\Psi^+_{AA'}\otimes \pi_{B'})U^\dagger_{A'B' \rightarrow CD}\right).
    \end{align}
    So then $I(A:C)_\omega=I(A:C)_{\rho^0}$, and so the second term in \eqref{eq:chainruleQAX} matches the second term in the averaged controllable correlation.
    It remains to show the first terms match, i.e. that $I(A:CX)_\omega=\sum_ip_i I(A:C)_{\rho^i}$. 
    For this, first use the chain rule
    \begin{align}
        I(A:CX)_{\omega}=I(A:C|X)_\omega+I(A:X)_\omega
    \end{align}
    and then observe that $A:X$ is uncorrelated, so the second term is zero. 
    Then, use lemma \ref{lem:averagedCMI} which gives
    \begin{align}
        I(A:CX)_{\omega}&=I(A:C|X)_\omega \nonumber \\
        &=\sum_{i}p_i I(A:C)_{\rho^i}
    \end{align}
    as claimed. 
\end{proof}

We prove the following relationship between the average and optimized controllable correlations. 
\begin{lemma}\label{lem:averageCCisnotlargerthanoptCC}
    Consider any ensemble $E=\{(p_i, \phi^i_{B'})\}$ with $\sum_i p_i \phi^i_{B'}=\pi_{B'}$, where $\pi_{B'}$ is the maximally mixed state. Then 
    \begin{align}\label{eq:avgvsopt}
        \overline{CC}(U,E)\leq CC^{\opt}(U). 
    \end{align}
\end{lemma}
\begin{proof}\,
    Choose $i^*$ where $\phi^{i^*}$ is the state among those in the ensemble $E$ that has the largest value of $I(A:C)_{\rho^{i^*}}$. 
    Then the controllable correlation $CC(U,\Psi^+, \phi^{i^*}, \pi)$ would be 
    \begin{align}
        CC(U, \Psi^+,\phi^{i^*}, \pi)&=I(A:C)_{\rho^{i^*}}-I(A:C)_{\rho^0} \nonumber \\
        &\geq I(A:C)_{\rho^{i}}-I(A:C)_{\rho^0}
    \end{align}
    which holds for every $i$. But then this also holds on average over $i$ with respect to the probabilities $p_i$, so 
    \begin{align}
        CC(U, \Psi^+,\phi^{i^*}, \pi) \geq \overline{CC}(U,E)
    \end{align}
    and finally $CC^{\opt}(U)\geq CC(U,\Psi^+, \phi^{i^*}, \pi)$, which proves \eqref{eq:avgvsopt}.   
\end{proof}

Note that the above lemma implies that the averaged controllable correlation also gives a lower bound on entanglement cost.

\subsection{Faithfulness of the controllable correlation}

To see that the controllable correlation is faithful, we prove the averaged controllable correlation being zero implies the scrambling value $J(U)$ is zero. 
In particular, this holds when the averaged controllable correlation is computed with respect to a \textbf{complete ensemble}, where we say that $E=\{(p_i,\phi^i_{B'})\}$ is a \textbf{complete ensemble} if $p_i>0$ for all $i$, $\sum_i p_i \phi^i_{B'}=\pi_{B'}$ and $\{\phi^i_{B'}\}$ spans the operators on $\mathcal{H}_{B'}$.

\begin{theorem}\label{thm:zeroaverageCC_implies_zeroscrambling}
    For every complete ensemble $E$, we have that
    \begin{align}
        \overline{CC}(U,E)=0\,\Rightarrow \, J(U)=0
    \end{align}
\end{theorem}
\begin{proof}\,
    First, assume that the averaged controllable correlation is zero. 
    From lemma \ref{lem:averageCC_CMI}, we have
    \begin{align}
        0=\overline{CC}(U,E)=I(A:X|C)_\omega.
    \end{align}
    This tells us, using the recovery theorem (theorem \ref{thm:equalityinSSA}), that 
    \begin{align}\label{eq:omegaform}
        \omega_{ACX}=\mathcal{T}_{C\rightarrow AC}(\omega_{CX}).
    \end{align}
    Recall that
    \begin{align}
        \omega_{ACX}=\sum_i p_i\rho^i_{AC}\otimes \ketbra{i}{i}_X.
    \end{align}
    Projecting in $\bra{i}_X\cdot \ket{i}_X$ to both sides of equation \eqref{eq:omegaform}, we get
    \begin{align}\label{eq:recovery}
        \rho^{i}_{AC}=\mathcal{T}_{C\rightarrow AC}(\rho_C^i),
    \end{align}
    where $\rho_C^i=\tr_A(\rho^{i}_{AC})$.
    Now define
    \begin{align}
        \mathcal{N}_{A'B' \rightarrow C}(\cdot)=\tr_D(U_{A'B'\rightarrow CD} (\cdot) U_{A'B'\rightarrow CD}^\dagger),
    \end{align}
    and use
    \begin{align}
        \rho^i_{AC}=\mathcal{N}_{A'B'\rightarrow C}(\Psi^+_{AA'}\otimes \phi^i_{B'}) = \mathcal{F}_{B'\rightarrow AC}(\phi^i_{B'})
    \end{align}
    where the last line defines $\mathcal{F}_{B'\rightarrow AC}$, and 
    \begin{align}
        \rho^i_{C}=\tr_{A}\rho_{AC}^i=\mathcal{N}_{A'B'\rightarrow C}(\pi_{A'}\otimes \phi^i_{B'}) = \mathcal{G}_{B'\rightarrow C}(\phi^i_{B'})
    \end{align}
    where the last equality defines $\mathcal{G}_{B'\rightarrow C}$. 
    Inserting the last two expressions in equation \eqref{eq:recovery}, we have
    \begin{align}
        \mathcal{F}_{B'\rightarrow AC}(\phi^i_{B'}) = \mathcal{T}_{C\rightarrow AC}\circ \mathcal{G}_{B'\rightarrow C}(\phi^i_{B'}).
    \end{align}
    Because the ensemble $E$ is complete, this holds for a spanning set of $\phi^i_{B'}$, and we get that 
    \begin{align}\label{eq:channelcomposition}
        \mathcal{F}_{B'\rightarrow AC}=\mathcal{T}_{C\rightarrow AC}\circ \mathcal{G}_{B'\rightarrow C}.
    \end{align}
    Next, let's look at the Choi state of $U_{A'B'\rightarrow CD}$, with $D$ traced out, 
    \begin{align}
        \chi_{ACB}^U &= \tr_{D}\left(U_{A'B'\rightarrow CD}(\Psi^{+}_{AA'}\otimes \Psi^+_{B'B}) U^{\dagger}_{A'B'\rightarrow CD} \right) \nonumber \\
        &=\mathcal{N}_{A'B'\rightarrow C}(\Psi^+_{AA'}\otimes \Psi^+_{B'B})\nonumber \\
        &=\mathcal{F}_{B'\rightarrow AC}(\Psi^+_{B'B})
    \end{align}
    where the last line follows from the definition of $\mathcal{F}_{B'\rightarrow AC}$.
    Now use equation \eqref{eq:channelcomposition}, leading to
    \begin{align}
        \chi_{ACB}^U &= \mathcal{T}_{C\rightarrow AC}\circ \mathcal{G}_{B'\rightarrow C}(\Psi^+_{B'B}) = \mathcal{T}_{C\rightarrow AC}(\chi^U_{CB}).
    \end{align}
    This shows that $\chi^U_{ACB}$ is a $B\rightarrow C\rightarrow A$ Markov chain, so then $J(U)=I(A:B|C)_{\chi^U}=0$, as claimed. 
\end{proof}

From here, we immediately obtain the following. 

\begin{theorem}
    For any complete ensemble $E$, the averaged controllable correlation is faithful. As well, the optimized controllable correlation is faithful. 
\end{theorem}
\begin{proof}\,
    To see that the averaged controllable correlation is faithful, use that from theorem \ref{thm:zeroaverageCC_implies_zeroscrambling} we get
    \begin{align}
        \overline{CC}(U,E)=0\Rightarrow J(U)=0
    \end{align}
    and from theorem \ref{thm:DingHaydenWalter}, $J(U)=0$ means $U$ is a subsystem SWAP unitary, so can be implemented with zero entanglement. 

    To see that the optimized controllable correlation is faithful we just use lemma \ref{lem:averageCCisnotlargerthanoptCC}, which states that
    \begin{align}
        CC^{\opt}(U)\geq \overline{CC}(U,E)
    \end{align}
    for any complete ensemble $E$, so when the optimized controllable correlation is zero, the average controllable correlation is zero, so then the unitary has the subsystem SWAP form and hence has zero cost. 
\end{proof}

\subsection{Achieving the maximal value}

The controllable correlation has the obvious upper bound
\begin{align}
    CC^{\opt}(U)= \max_{P_{AA'}, \phi_{B'}^1, \phi^2_{B'}}\left( I(A:C)_{\rho^1}-I(A:C)_{\rho^2}\right) \leq 2n_A.
\end{align}
In \cite{cleve2026lower}, the controllable correlation was computed for a variety of two qubit unitaries. 
It was observed that the largest obtained value appeared to be $1$, even when sampling a large number of Haar random two qubit unitaries, which is suggestive that the controllable correlation actually attains a maximum at $n_A$ (since $n_A=1$ for a two qubit unitary).   

However, we can actually saturate the bound of $2n_A$ easily. 
Consider the controlled SWAP unitary, 
\begin{align}\label{eq:controlledSWAP}
    U_{A'B'\rightarrow CD} = \mathcal{I}_{A'B'_1} \otimes \ketbra{0}{0}_{B'_0}+ \SWAP_{A'B'_1}\otimes \ketbra{1}{1}_{B'_0}.
\end{align}
Here $B'=B_0'B_1'$, and we identify $A'\rightarrow C$, $B'\rightarrow D$ in the outputs.  
Choose $\phi^1_{B'}=\ketbra{0}{0}_{B_0'}\otimes \pi_{B_1'}$, $\phi^2_{B'}=\ketbra{1}{1}_{B_0'}\otimes \pi_{B_1'}$, and $P_{AA'}=\Psi^+_{AA'}$. 
Then we find that
\begin{align}
    \lambda_1&=I(A:C)_{\rho^1} = 2n_A, \nonumber \\
    \lambda_2&=I(A:C)_{\rho^2} = 0
\end{align}
so that
\begin{align}
    CC^{\opt}(U)= 2n_A. 
\end{align}

\section{Relationship between controllable correlation and the scrambling quantity}\label{sec:relationship_CC_scrambling}

So far, we have studied three quantities which provide lower bounds on entanglement cost in NLQC: the optimized controllable correlation $CC^{\opt}(U)$, the averaged controllable correlation $\overline{CC}(U,E)$, and the scrambling quantity $J(U)$. 
We noted already that
\begin{align}
    CC^{\opt}(U)\geq \overline{CC}(U,E).
\end{align}
It is natural to ask about the remaining relationships among these three quantities. 
We first of all clarify the relationship between the averaged controllable correlation and the scrambling quantity. 

\begin{remark}\label{remark:JisnotlessthanaverageCC}
    For every ensemble $E=\{(p_i, \phi^i)\}$ such that $\sum_i p_i \phi^i=\pi$, the averaged controllable correlation and the scrambling quantity are related by
    \begin{align}
        J(U)\geq \overline{CC}(U,E)
    \end{align}
\end{remark}
\begin{proof}\,
    Recall that the scrambling quantity is
    \begin{align}
        J(U) = I(A:B|C)_{\chi^U}
    \end{align}
    where $\chi^U$ is the Choi state of $U$, 
    \begin{align}
        \chi^U_{ABCD} = U_{A'B'\rightarrow CD}\left(\Psi^{+}_{AA'}\otimes \Psi^+_{BB'} \right)U^\dagger_{A'B'\rightarrow CD}.
    \end{align}
    Meanwhile by lemma \ref{lem:averageCC_CMI}
    \begin{align}
        \overline{CC}(U,E)=I(A:X|C)_{\omega}
    \end{align}
    where
    \begin{align}
        \omega_{ACDX} = U_{A'B'\rightarrow CD}\left(\Psi^{+}_{AA'}\otimes \sum_i p_i\phi^i_{B'}\otimes \ketbra{i}{i}_X\right)U^\dagger_{A'B'\rightarrow CD}.
    \end{align}
    Thus $J(U)$ and $\overline{CC}(U,E)$ are very similar quantities. 
    To relate them, consider beginning with $J(U)$ and measuring the $B$ system with the POVM $\{\Lambda^i_{B}=p_id_{B}(\phi^i_{B})^T\}_i$. 
    We can check this is a legitimate POVM, since the $\Lambda^i$ are positive since the $\phi^i$ are, and $\sum_i \Lambda^i=\mathcal{I}$ because $\sum_i p_i \phi^i=\pi$.
    Consider measuring the $B$ subsystem of ${\Psi^+}_{B'B}$ using this POVM, and recording the measurement outcome $i$ in the $X$ register, 
    \begin{align}
        \mathcal{M}_{B\rightarrow X}(\Psi^+_{B'B}) &= \sum_i \tr_B(\Lambda^i_B \Psi^+_{B'B})\otimes \ketbra{i}{i}_X \nonumber \\
        &= \sum_i p_i\phi^i_{B'}\otimes \ketbra{i}{i}_X.
    \end{align}
    We get that this is exactly the input state when computing the controllable correlation. 
    But by the data processing inequality (lemma \ref{lem:dataprocessinginequality}), we have that
    \begin{align}
        J(U) = I(A:B|C)_{\chi^U}\geq I(A:X|C)_{\mathcal{M}_{B\rightarrow X}(\chi^U)} = I(A:X|C)_{\omega} = \overline{CC}(U,E)
    \end{align}
    as claimed. 
\end{proof}

We see that the averaged controllable correlation is bounded above by both the optimized controllable correlation and the scrambling quantity, so provides the weakest lower bound. 
It remains to understand how $J(U)$ and $CC^{\opt}(U)$ are related. 
In fact, we can observe that they are incomparable, with neither quantity always providing the better bound. 
Instead, the two quantities provide complementary bounds. 

To see this is straightforward: we can simply evaluate both bounds on the unitaries we've constructed that maximize each bound. 

First, consider the controlled SWAP unitary (equation \eqref{eq:controlledSWAP}), which we observed obtains a maximal value for the controllable correlation of $2n_A$. 
For concreteness choose $n_A=1$, so that $n_B=2$. 
We found that $CC^{\opt}(\mathrm{CSWAP})=2$. 
Meanwhile, we can straightforwardly evaluate the scrambling quantity, and we find that $J(\mathrm{CSWAP})\approx 0.55$. 
Thus the optimized controllable correlation can be larger than the scrambling quantity. 

Next, lets revisit the explicit maximizing unitary we found for the scrambling quantity, the unitary in equation \eqref{eq:perfectunitary}. 
For concreteness choose $d_A=d_B=3$. 
This achieves the maximal value of the scrambling quantity, $J(U^{\text{perfect}})=2\log 3$.
Meanwhile, a numerical evaluation of the optimized controllable correlation returns $1.58$ (likely in fact $\log 3$).
Thus the scrambling quantity can be larger than the controllable correlation.

\section{Discussion}\label{sec:discussion}

In this work we give a new lower bound on entanglement cost in NLQC. 
Our bound is based on the scrambling quantity, which for unitary channels reduces to (minus) the tripartite mutual information used as a diagnostic of information scrambling. 
For unitaries, our bound is faithful: the bound never fails to return a positive entanglement cost if the unitary really does require entanglement. 
This is the first faithful lower bound for unitaries, and the relationship between scrambling and NLQC is new. 
Finally, the scrambling quantity can be shown to be zero whenever the controllable correlation is zero, showing that this earlier lower bound is also faithful.

These results begin to illuminate the origin of entanglement cost in NLQC. 
However, a complete understanding of entanglement cost in NLQC remains far away. 
Indeed, as noted elsewhere, many questions about entanglement cost in NLQC relate to difficult open problems in complexity theory, and we do not expect to understand these problems in full in the near future. 
Even many simple questions about entanglement cost in NLQC remain open however. 
For instance, we now have two faithful lower bounds for entanglement cost in NLQC, and one of them (the scrambling quantity) can be evaluated for any channel. 
However, the scrambling bound is not faithful for general quantum channels. 
For example, let $U^\text{perfect}$ be the unitary in \eqref{eq:perfectunitary} with $d=3$ and let $\mathcal{R}(\cdot)=\ketbra{0}{0}$ be the reset channel. Consider $\mathcal{N}=U^\text{perfect}\otimes \mathcal{R}$ where $\mathcal{R}$ acts on an additional qutrit held by Alice. Then, $J(\mathcal{N})=0$ and $E(\mathcal{N})\geq E(U^\text{perfect})\geq \log 3$.
An interesting open question is to find a faithful lower bound for quantum channels. 
More generally, an understanding of the general, qualitative, properties that lead to entanglement cost in channels is still missing --- the controllable correlation bound fails to hold in that setting, and the `scrambling quantity' defined on general channels is not clearly related to the scrambling properties of the channel\footnote{Recall that it is clearly related to scrambling for unitaries, and we keep the naming in the case of general channels for convenience.} or more generally to have a simple meaning.

Addressing generic approaches to bounding entanglement cost in quantum channels may be an interesting, alternative, approach to bounding entanglement cost in the specific channels of the strongest interest. 
For instance in the $f$-routing scenario \cite{buhrman2013garden, kent2011quantum, cree2023code, bluhm2026complexity, bluhm2026equivalence}, a qubit is redirected left or right based on a Boolean function $f$ of classical inputs $x,y\in\{0,1\}^n$. 
An important goal is to prove a lower bound on entanglement cost that grows super-logarithmically with $n$, even while the size of the quantum system is kept fixed. 
Several partial results in this direction have failed to give fully satisfactory bounds \cite{asadi2025rank, bluhm2021position, asadi2025linear, may2026quantum}.
The scrambling bound given here also fails to give strong entanglement bounds, but we can hope that further understanding of generic properties of channels that lead to entanglement cost can shed light on this problem. 

\vspace{0.2cm}
\noindent \textbf{Acknowledgements:} A. H. is supported by JSPS KAKENHI grants Nos.~24H00071, 25K24674, 25K24465. A. M. acknowledges the support of the Natural Sciences and Engineering Research Council of Canada (NSERC); this work was supported by an NSERC-UKRI Alliance Grant No. (ALLRP 597823-24). 
Research at Perimeter Institute is supported in part by the Government of Canada through the Department of Innovation, Science and Economic Development and by the Province of Ontario through the Ministry of Colleges and Universities.
 
OpenAI Astra and Claude Opus were used extensively to suggest and simplify proofs. 
Astra initially gave a proof that controllable correlation is faithful. 
After understanding the proof, which revealed the subsystem SWAP unitaries as exactly the zero entanglement unitaries, we suggested Astra look for a more direct proof that these are the zero entanglement unitaries. 
Astra then returned with a proof of that claim based on theorem \ref{thm:DingHaydenWalter}. 
We then observed that another lower bound on entanglement based on the scrambling quantity was being implied by this new proof, and we refined this using help from Claude Opus to develop theorem \ref{thm:Jlowerbound}.
Further refinements, including the extension of the scrambling bound to general channels, were then done using Claude Opus.

\bibliographystyle{unsrtnat}
\bibliography{biblio}

@article{may2022complexity,
  doi = {10.22331/q-2022-11-28-864},
  opt-url = {10.22331/q-2022-11-28-864},
  title = {Complexity and entanglement in non-local computation and holography},
  author = {May, Alex},
  journal = {{Quantum}},
  opt-issn = {2521-327X},
  publisher = {{Verein zur F{\"{o}}rderung des Open Access Publizierens in den Quantenwissenschaften}},
  volume = {6},
  pages = {864},
  opt-month = nov,
  year = {2022}
}

@article{bluhm2021position,
  title={A single-qubit position verification protocol that is secure against multi-qubit attacks},
  author={Bluhm, Andreas and Christandl, Matthias and Speelman, Florian},
  journal={Nature Physics},
  pages={1--4},
  year={2022},
  publisher={Nature Publishing Group},
  doi={10.1038/s41567-022-01577-0}
}

@inproceedings{buhrman2013garden,
  title={The garden-hose model},
  author={Buhrman, Harry and Fehr, Serge and Schaffner, Christian and Speelman, Florian},
  booktitle={Proceedings of the 4th conference on Innovations in Theoretical Computer Science (ITCS 2013)},
  pages={145--158},
  year={2013},
  doi={10.1145/2422436.2422455}
}

@article{may2020holographic,
  title={Holographic scattering requires a connected entanglement wedge},
  author={May, Alex and Penington, Geoff and Sorce, Jonathan},
  journal={Journal of High Energy Physics},
  volume={2020},
  number={8},
  pages={1--34},
  year={2020},
  publisher={Springer},
  doi={10.1007/JHEP08(2020)132}
}

@article{may2019quantum,
  title={Quantum tasks in holography},
  author={May, Alex},
  journal={Journal of High Energy Physics},
  volume={2019},
  number={10},
  pages={1--39},
  year={2019},
  publisher={Springer},
  doi={10.1007/JHEP10(2019)233}
}

@article{tomamichel2013monogamy,
  title={A monogamy-of-entanglement game with applications to device-independent quantum cryptography},
  author={Tomamichel, Marco and Fehr, Serge and Kaniewski, J{\k{e}}drzej and Wehner, Stephanie},
  journal={New Journal of Physics},
  volume={15},
  number={10},
  pages={103002},
  year={2013},
  publisher={IOP Publishing},
  doi={10.1088/1367-2630/15/10/103002}
}

@article{junge2021geometry,
  title={Geometry of {B}anach spaces: a new route towards {P}osition {B}ased {C}ryptography},
  author={Junge, Marius and Kubicki, Aleksander M and Palazuelos, Carlos and P{\'e}rez-Garc{\'\i}a, David},
  journal={arXiv preprint arXiv:2103.16357},
  year={2021},
  doi={10.1007/s00220-022-04407-9}
}

@InProceedings{speelman2015instantaneous,
  author =	{Florian Speelman},
  title =	{{Instantaneous Non-Local Computation of Low T-Depth Quantum Circuits}},
  booktitle =	{11th Conference on the Theory of Quantum Computation, Communication and Cryptography (TQC 2016)},
  pages =	{9:1--9:24},
  series =	{LIPIcs},
  opt-ISBN =	{978-3-95977-019-4},
  ISSN =	{1868-8969},
  year =	{2016},
  volume =	{61},
  opt-publisher =	{Schloss Dagstuhl--Leibniz-Zentrum fuer Informatik},
  opt-address =	{Dagstuhl, Germany},
  URN =		{urn:nbn:de:0030-drops-66909},
  doi =		{10.4230/LIPIcs.TQC.2016.9}
}

@article{kent2011quantum,
  title={Quantum tagging: Authenticating location via quantum information and relativistic signaling constraints},
  author={Kent, Adrian and Munro, William J and Spiller, Timothy P},
  journal={Physical Review A},
  volume={84},
  number={1},
  pages={012326},
  year={2011},
  publisher={APS},
  doi={10.1103/PhysRevA.84.012326}
}

@article{buhrman2014position,
  title={Position-based quantum cryptography: Impossibility and constructions},
  author={Buhrman, Harry and Chandran, Nishanth and Fehr, Serge and Gelles, Ran and Goyal, Vipul and Ostrovsky, Rafail and Schaffner, Christian},
  journal={SIAM Journal on Computing},
  volume={43},
  number={1},
  pages={150--178},
  year={2014},
  publisher={SIAM},
  doi={10.1137/130913687}
}

@article{beigi2011simplified,
  title={Simplified instantaneous non-local quantum computation with applications to position-based cryptography},
  author={Beigi, Salman and K{\"o}nig, Robert},
  journal={New Journal of Physics},
  volume={13},
  number={9},
  pages={093036},
  year={2011},
  publisher={IOP Publishing},
  doi={10.1088/1367-2630/13/9/093036}
}

@article{cree2023code,
  title={Code-routing: a new attack on position verification},
  author={Cree, Joy and May, Alex},
  journal={Quantum},
  volume={7},
  pages={1079},
  year={2023},
  publisher={Verein zur F{\"o}rderung des Open Access Publizierens in den Quantenwissenschaften}
}

@article{gonzales2019bounds,
  title={Bounds on instantaneous nonlocal quantum computation},
  author={Gonzales, Alvin and Chitambar, Eric},
  journal={IEEE Transactions on Information Theory},
  volume={66},
  number={5},
  pages={2951--2963},
  year={2019},
  publisher={IEEE},
  doi={10.1109/TIT.2019.2950190}
}

@article{apel2024security,
  title={Security of position-based quantum cryptography limits {H}amiltonian simulation via holography},
  author={Apel, Harriet and Cubitt, Toby and Hayden, Patrick and Kohler, Tamara and P{\'e}rez-Garc{\'\i}a, David},
  journal={arXiv preprint arXiv:2401.09058},
  year={2024},
doi={10.48550/arXiv.2401.09058}
}

@article{allerstorfer2023relating,
  title={Relating non-local quantum computation to information theoretic cryptography},
  author={Allerstorfer, Rene and Buhrman, Harry and May, Alex and Speelman, Florian and Lunel, Philip Verduyn},
  journal={Quantum},
  volume={8},
  pages={1387},
  year={2024},
  publisher={Verein zur F{\"o}rderung des Open Access Publizierens in den Quantenwissenschaften},
  doi = {10.22331/q-2024-06-27-1387}
}

@inproceedings{ananth2025unclonable,
  title={Unclonable Secret Sharing},
  author={Ananth, Prabhanjan and Goyal, Vipul and Liu, Jiahui and Liu, Qipeng},
  booktitle={International Conference on the Theory and Application of Cryptology and Information Security},
  pages={129--157},
  year={2025},
  organization={Springer}
}

@article{may2026entanglement,
  title={Entanglement cost in non-local quantum computation},
  author={May, Alex},
  journal={arXiv preprint arXiv:2605.02840},
  year={2026},
  doi = {10.48550/arXiv.2605.02840}
}

@InProceedings{asadi2025rank,
  author =	{Asadi, Vahid R. and Culf, Eric and May, Alex},
  title =	{{Rank Lower Bounds on Non-Local Quantum Computation}},
  booktitle =	{16th Innovations in Theoretical Computer Science Conference (ITCS 2025)},
  pages =	{11:1--11:18},
  series =	{LIPIcs},
  ISSN =	{1868-8969},
  year =	{2025},
  volume =	{325},
  opt-address =	{Dagstuhl, Germany},
  URN =		{urn:nbn:de:0030-drops-226399},
  doi =		{10.4230/LIPIcs.ITCS.2025.11}
}

@article{asadi2025linear,
  title={Linear gate bounds against natural functions for position-verification},
  author={Asadi, Vahid and Cleve, Richard and Culf, Eric and May, Alex},
  journal={Quantum},
  volume={9},
  pages={1604},
  year={2025},
  publisher={Verein zur F{\"o}rderung des Open Access Publizierens in den Quantenwissenschaften},
  doi = {10.22331/q-2025-01-21-1604}
}

@article{bluhm2026complexity,
  title={A complexity theory for non-local quantum computation},
  author={Bluhm, Andreas and H{\"o}fer, Simon and May, Alex and Stasiuk, Mikka and Lunel, Philip Verduyn and Yuen, Henry},
  journal={Quantum},
  volume={10},
  pages={2152},
  year={2026},
  publisher={Verein zur F{\"o}rderung des Open Access Publizierens in den Quantenwissenschaften},
  doi = {10.22331/q-2026-07-01-2152}
}

@article{bluhm2026equivalence,
  title={Equivalence of non-local computation tasks beyond Clifford operations},
  author={Bluhm, Andreas and H{\"o}fer, Simon and May, Alex and Speelman, Florian and Lunel, Philip Verduyn},
  journal={arXiv preprint arXiv:2606.26354},
  year={2026},
  doi = {10.48550/arXiv.2606.26354}
}

@article{cleve2026lower,
  title={Lower bounds on non-local computation from controllable correlation},
  author={Cleve, Richard and May, Alex},
  journal={arXiv preprint arXiv:2602.00255},
  year={2026}
}

@article{hayden2004structure,
  title={Structure of states which satisfy strong subadditivity of quantum entropy with equality},
  author={Hayden, Patrick and Jozsa, Richard and Petz, Denes and Winter, Andreas},
  journal={Communications in Mathematical Physics},
  volume={246},
  number={2},
  pages={359--374},
  year={2004},
  publisher={Springer}
}

@article{ding2016conditional,
  title={Conditional mutual information of bipartite unitaries and scrambling},
  author={Ding, Dawei and Hayden, Patrick and Walter, Michael},
  journal={Journal of High Energy Physics},
  volume={2016},
  number={12},
  pages={1--32},
  year={2016},
  publisher={Springer}
}

@article{asadi2025conditional,
  title={Conditional disclosure of secrets with quantum resources},
  author={Asadi, Vahid R and Kuroiwa, Kohdai and Leung, Debbie and May, Alex and Pasterski, Sabrina and Waddell, Chris},
  journal={Quantum},
  volume={9},
  pages={1885},
  year={2025},
  publisher={Verein zur F{\"o}rderung des Open Access Publizierens in den Quantenwissenschaften}
}

@inproceedings{girish2026magic,
  title={Magic and communication complexity},
  author={Girish, Uma and May, Alex and Parham, Natalie and Yuen, Henry},
  booktitle={Proceedings of the 58th Annual ACM Symposium on Theory of Computing (STOC 2026)},
  pages={845--856},
  year={2026}
}

@article{moreno2026exponential,
  title = {Exponential lower bounds for low-degree strategies in position-based quantum cryptography},
  author = {Moreno-Cuadrado, Isabel M and Perez-Garcia, David and Palazuelos, Carlos},
  journal = {arXiv preprint arXiv:2609.35592},
  year = {2026},
}

@article{may2024non,
  title={Non-local computation and the black hole interior},
  author={May, Alex and Xu, Michelle},
  journal={Journal of High Energy Physics},
  volume={2024},
  number={2},
  pages={79},
  year={2024},
  publisher={Springer}
}

@article{may2022connected,
  title = {The connected wedge theorem and its consequences},
  author = {May, Alex and Sorce, Jonathan and Yoshida, Beni},
  journal = {Journal of High Energy Physics},
  volume = {2022},
  number = {11},
  pages = {1--65},
  year = {2022},
  publisher = {Springer},
}

@article{hayden2013holographic,
  title={Holographic mutual information is monogamous},
  author={Hayden, Patrick and Headrick, Matthew and Maloney, Alexander},
  journal={Physical Review D—Particles, Fields, Gravitation, and Cosmology},
  volume={87},
  number={4},
  pages={046003},
  year={2013},
  publisher={APS}
}

@article{hosur2016chaos,
  title={Chaos in quantum channels},
  author={Hosur, Pavan and Qi, Xiao-Liang and Roberts, Daniel A and Yoshida, Beni},
  journal={Journal of High Energy Physics},
  volume={2016},
  number={2},
  pages={4},
  year={2016},
  publisher={Springer}
}

@article{PRXQuantum.5.010201,
  title = {Scrambling Dynamics and Out-of-Time-Ordered Correlators in Quantum Many-Body Systems},
  author = {Xu, Shenglong and Swingle, Brian},
  journal = {PRX Quantum},
  volume = {5},
  issue = {1},
  pages = {010201},
  numpages = {46},
  year = {2024},
  opt-month = {Jan},
  publisher = {American Physical Society},
  doi = {10.1103/PRXQuantum.5.010201},
  opt-url = {https://link.aps.org/doi/10.1103/PRXQuantum.5.010201},
}

@article{may2026quantum,
  title = {Quantum gate lower bounds for loss-tolerant position verification},
  author = {May, Alex and Lunel, Philip Verduyn},
  journal = {arXiv preprint arXiv:2608.16495},
  year = {2026},
}

\end{document}